\documentstyle[12pt,aaspp4]{article}
\newcommand{\be}{\begin{equation}}
\newcommand{\ee}{\end{equation}}
\newcommand{\ba}{\begin{eqnarray}}
\newcommand{\ea}{\end{eqnarray}}
\begin{document}
\title{Inelastic Dissipation in a Freely Rotating Body.\\ 
Application to Cosmic-Dust Alignment.}
\author{A. Lazarian\altaffilmark{1}}
\affil{Princeton University Observatory, Princeton NJ 08544 USA}
%\address{Princeton University Observatory, Princeton NJ 08544 USA}
\author{Michael Efroimsky\altaffilmark{2}}
%\address{Institute of Theoretical Atomic and Molecular Physics\\
%and\\
%Department of Physics, Harvard University,\\ 
%17 A Oxford Street, Cambridge MA 02138 USA} 
\affil{Institute of Theoretical Atomic and Molecular Physics\\
and\\
Department of Physics, Harvard University,\\ 
17 A Oxford Street, Cambridge MA 02138 USA}
%\date{\today}
%\maketitle
\begin{abstract}
Motivated by a recent study by Lazarian and Draine, 
 which showed that a high  degree of grain alignment of the paramagnetic 
dust is achievable if the rates of internal relaxation are 
controlled by the Barnett relaxation process, 
we undertake a study of an alternative mechanism of
internal dissipation, namely, the inelastic dissipation of energy
in oblate dust grains. We find that deformations at double
frequency that were disregarded in earlier studies dominate the inelastic
relaxation. Our results indicate that for sufficiently prolate and oblate
grains, e.g. for grains with 4:1 axis ratio, or/and grains formed
by agglomeration
inelastic relaxation dominates the Barnett relaxation within
large ($a>0.1$~$\mu$m) grains. For grains with axis ratio less than
1:2 the inelastic relaxation is dominant for suprathermally rotating
grains.
\end{abstract}


\section{Introduction}

The problem of inelastic dissipation of a freely rotating body emerges 
naturally in the context of interstellar grain alignment.
 
It is well known that the polarization of starlight is caused by 
the alignment 
of interstellar dust grains. This was discovered by Hiltner and Hall 
(see Hall 1949  and references therein) who showed that the polarization 
was entailed by the linear dichroism, 
i.e., by the differential extinction, by clouds of aligned grains, of 
electromagnetic waves of different polarizations.

Various mechanisms of grain alignment (see table~1 in Lazarian, Goodman
\& Myers 1997) depend on the internal dissipation of energy (Purcell 1979).
This dissipation aligned the axis of the maximal moment of inertia
and the angular momentum. This changes grain dynamics and should be
accounted for in the theory of grain alignment. Therefore the  
grain alignment depends on the internal relaxation. 
In some cases, e.g. the alignment of thermally
rotating grains, the alignment is not sensitive to the exact rate of
internal relaxation. However for grains rotating suprathermally, 
more detailed knowledge of the internal relaxation is usually
required (see more
in the discussion section of this paper). 

It is normally believed that the Barnett dissipation is the dominant 
process of internal relaxation within interstellar grains (Purcell 1979,
Spitzer \& McGlynn 1979, Lazarian 1994, Draine 1996). 
It happens due to the time-dependent (due to the 
precession of angular velocity in grain body axes) remagnetisation of 
the sample by means of the Barnett effect. 
Indeed, the Barnett effect develops magnetization parallel to the angular
velocity of a rotating body because a share of the angular momentum gets 
transferred to the spins (see Landau and Lifshitz 1984). In the cause of the 
angular-velocity precession about the angular momentum (in accordance with 
the Euler equations), the 
magnetization direction is also precessing in the body-axes frame, and this 
entails dissipation. A recent study of paramagnetic alignment of suprathermally
rotating grains (or ``Purcell alignment'') has shown that grains can 
be nearly perfectly aligned by the interstellar magnetic field if the
internal dissipation is controlled by the Barnett effect (Lazarian
\& Draine 1997).

Another process responsible for dissipation, known as inelastic 
relaxation happens as time-dependent stresses emerge in a body when 
it rotates about an arbitrary axis. In this paper we subject this process 
to scrutiny, as the existing treatment (Purcell 1979) has a number of 
difficiencies that we discuss in the paper. 

A rigorous treatment of internal relaxation is important: if the inelastic 
relaxation is more efficient than the Barnett relaxation then it is the 
former, not the latter, that determines the dynamics of crossovers 
and, thereby, the degree of alignment achievable by suprathermally
rotating grains (see Lazarian \& Draine 1997).

In what follows we briefly discuss the rotation of an oblate  
grain (Section~II); calculate the acceleration experienced by a point inside 
the grain (Section~III). Then we compute the stresses caused by grain 
precession (Section~IV), the rate of energy dissipation (Section~V) and the 
rate of internal-dissipation-caused alignment of the major-inertia axis of 
the grain to its angular momentum. We also discuss whether the obtained 
formulae could be easily altered in the case of prolate grains (Section~VI). 
Our results are discussed in Section~VII.

\section{Notations and Assumptions}

We consider a freely rotating cosmic-dust grain, 
employing two Cartesian coordinate systems, each with its origin at the 
center of mass of the body. The inertial frame ($X$, $Y$, $Z$), with 
unit vectors $\mathbf{e}$$_{X}$, $\mathbf{e}$$_{Y}$, $\mathbf{e}$$_{Z}$, 
is chosen so that its $Z$ axis is parallel to the (conserved) angular 
momentum $\mathbf{J}$ (see Fig.~1). Coordinates with respect to this frame are 
denoted by the same capital letters: $X$, $Y$, and $Z$. 
We also use the system which we call body frame 
and associate with the three principal axes of inertia: $1$, $2$, and 
$3$, with coordinates $x$, $y$, and $z$ along these. The appropriate unit 
vectors are $\mathbf{e}$$_{1}$, $\mathbf{e}$$_{2}$, 
$\mathbf{e}$$_{3}$.

Without loss of generality, one may take $I_3 \geq  I_1$ and $I_3 \geq I_2$,
where $I_i$ are the principal moments of inertia.

Following Purcell (1979), 
we denote the angular velocity by $\bf{\Omega}$ and reserve 
$\bf{\omega}$ for the rate of precession. 
It is a well-known fact that, as a perfectly elastic body rotates, the 
extremity of the angular velocity vector $\bf{\Omega}$ describes a curve that 
is the intersection of two ellipsoids. One of those comes from the 
angular-momentum conservation:
\be
I_1^2 \; \Omega_1^2 \; + \; I_2^2 \; \Omega_2^2 \; + 
\; I_3^2 \; \Omega_3^2 \; = \; {\bf{J}}^2 \; = \; const~~~,
\label{2.4}
\ee
where ${\Omega}_{1,2,3}$ are the body-frame-related components of 
$\bf{\Omega}$). Another one, known as the Poinsot ellipsoid, is 
defined by the kinetic-energy conservation:
\be
I_1 \; \Omega_1^2 \; + \; I_2 \; \Omega_2^2 \; + \; I_3 \; \Omega_3^2 \;
= \; {\bf{J \; \Omega}} \; = 2 \: E_{rot} \; = \; const~~~.
\label{2.5}
\ee
In the presence of inner dissipation, the 
Poinsot ellipsoid will be distorting in the cause of time, and will eventually 
degenerate into a line of length $I_3^{-1/2}$ oriented along the 
axis of maximal moment of inertia. That 
will happen after the body aligned its angular velocity along its 
maximal-inertia axis: ${\Omega}_1 = {\Omega}_2 = 0, \; |{\Omega}_3| = 
|{\bf{J}}|/I_3$. This mode minimizes the kinetic energy, with 
the angular momentum $\bf{J}$ fixed.

In what follows we  model the grain by an oblate symmetric top, e.g. a disk
\be
I_{3} \; > \; I_{1} \; = \; I_{2} \; \equiv \; I \; \; . 
\label{2.6}
\ee
In the grain-alignment studies such an approximation is customary (Purcell 
1979, Spitzer \& McGlynn 1979). For such a body the Euler equations read:
\be
\dot{\Omega}_1 \; + \: {\frac{I_3 - I}{I}} \; {{\Omega}_2} \; 
{{\Omega}_3} \; = \; 0 ~~~,
\label{6}
\ee
\be
\dot{\Omega}_2 \; + \: {\frac{I - I_3}{I}} \; {{\Omega}_3} \;
{{\Omega}_1} \; = \; 0 ~~~,
\label{7}
\ee
\be
\dot{\Omega}_3 \; = \; 0~~~, 
\label{8}
\ee
where ${\Omega}_i$ is the body-frame components of the angular velocity. 
As well known, the angular velocity ${\bf{\Omega}}$ nutates about 
the principal axis 3 at a  constant rate 
\be
\omega \; = \; (h - 1) \Omega_3, 
\; \; \; \; \; \; \; \;  h\equiv {I_3}/I~~~. 
\label{9}
\ee
However from the point of view of the inertial observer it is 
rather the principal axis 3 that is precessing 
about $\bf{J}$ (which is conserved in the inertial frame). Therefore 
angle $\theta$ is constant. 
Now, let $\alpha$ be the angle described by the precession of vector 
${\bf \Omega}$ about axis $3$ (or of axis $3$ about ${\bf \Omega}$). 
It is easy to see that 
\be
\tan {\theta} \; = \; 
\; (I \; {|{\bf{\Omega}}|} \; \; \sin \; {\alpha})/(I_3 \; 
{|{\bf{\Omega}}|} \; \; \cos \; {\alpha}) \; = \; h \;\tan {\alpha}~~~;
\label{fnote}
\ee
hence  $\alpha$ is also constant. As $\Omega_3 = \Omega\;\cos\alpha$, 
the angular velocity of precession given by (\ref{9}), 
 except a special case of ${\bf{\Omega}}$ and 
${\bf{J}}$  orthogonal or almost orthogonal to the maximal-inertia axis 
$3$, is typically of the order of $|{\bf{\Omega}}|$. 
Hence one may call the rotation and precession 
``fast motions'', implying that the relaxation is  a slow 
process (which means that the rate of alignment: $\; \dot{\theta}\ll\omega$). 

Before pursuing with the calculational part of our article, let us once 
again dwell upon the physical mechanism that changes $\Omega_3$ 
over the spans of time comparable with the internal relaxation time. 
In terms of energy, everything is clear: the  inelastic body 
aligns so as to minimize its rotational energy and still to conserve its 
angular momentum. This obvious mechanism does not seem to be instilled in the 
Euler equations ($\ref{6}$), ($\ref{7}$), ($\ref{8}$), for the torques remain 
to be zero, and nothing seems to be able to shift the solutions 
${{\Omega}_{1,2,3}}(t)$ from the form that yields $\alpha$ constant. 
In reality, 
however, this would-be contradiction is easily resolved if one recalls that 
inelastic relaxation stems from the inner displacements of grain material. 
These displacements result in changes of grain moments of inertia. Therefore
in the cause of the grain's wobbling, its principle moments of inertia 
fluctuate, and terms $\Omega_i\dot{I}_i \not = 0$ should be 
accounted for in Euler equations. Inelasticity results in a phase 
shift between the angular velocity and $\dot{I}_i$ and this 
causes internal alignment\footnote{The latter statement is equivalent
to the statement that there is a phase shift between the stresses
that act upon the grain, and the deformation of the grain.}. 

Since $\dot{I}_i$ depend on the grain elasticity, while the phase 
offset of $\dot{I}_i$ and $\Omega$ is determined by inelastic effects, then 
the parameters that determine elasticity and  dissipation within
the material should enter our
final formulae for the inelastic relaxation.

\section{Acceleration of a Point inside an Oblate Grain 
$ \; \; ( \; \; \bf{I_3 \; \geq \; I_1 \; = \; I_2 \; \; })$}

Our goal is to calculate $\dot{\theta}$, the rate of the 
maximal-inertia axis' approaching the direction of angular momentum $\bf{J}$. 
 To achieve this goal, one has to know the rate of energy losses entailed 
by the inelastic deformation. To calculate the deformation, 
we shall have to know 
the acceleration experienced by a particle located inside the grain at a 
point ($x$, $y$, $z$). Note that we address the proper acceleration, i.e.
that with respect to the inertial frame $(X,Y,Z)$, but we shall express it in 
terms of coordinates $x$, $y$ and $z$ of the body frame $(1,2,3)$. 
The fast processes (revolution and precession of a symmetric 
oblate grain) are described by the Euler equations ($\ref{6}$ - 
$\ref{8}$)  whose solution, in neglect of the slow relaxation, will read
\be
{\Omega}_1 \; \; = \; \; {\Omega}_{\perp} \cos {\omega}t~~,~~~
{\Omega}_2 \; \; = \; \; {\Omega}_{\perp} \sin {\omega}t~~,~~~
{\Omega}_3 \; \; = \; \; const
\label{rotation}
\ee
where
\be
{{\Omega}_{\perp}} \; \; = \; \; {\Omega} \; \; \sin \; {\alpha}~~, ~~~~~
{{\Omega}_{3}} \; \; = \; \; {\Omega} \; \; \cos \; {\alpha}
\label{24}  
\ee
and
\be
{{\Omega}_{\perp}}/{{\Omega}_{3}} \; \; = \; \; \tan \; \alpha \; \; = \; \; 
h \; \; \tan \; {\theta}~~~.
\label{25}
\ee

Besides, we shall need formulae connecting the components of $\bf{\Omega}$ 
with the absolute values of the angular momentum:
\be
{\Omega}_3 \; \; = \; \; \frac{J_3}{I_3} \; \; = \; \; \frac{J}{I_3} \; \; 
\cos \; \theta~~, ~~~~
{\Omega}_{\perp} \; \; = \; \; \frac{J}{I_3} \; h \; \; \sin \; \theta
\label{om4}
\ee
We denote the position, velocity and acceleration relative to the 
body frame as: ${\bf{r}}$, ${\bf{v}}$, ${\bf{a} \;}$, while those
related to the body frame $ \; (1,2,3) \; $ will be called ${\bf{r''}}$, 
${\bf{v''}}$, ${\bf{a''} \;}$, where we shall keep in mind that $\;{\bf{r}} = 
{\bf{r''}}$. 
The acceleration ${\bf{a} \;}$ in the inertial frame looks:
\be
{\bf{a}} \; \; = \; \; {\bf{a''}} \; \; + \; \; 
{\bf{\dot{\Omega}}} \; \times \; {\bf{r''}} \; \; + \; \; 
2 \; \; {\bf{\Omega}} \; \times \; {\bf{v''}} \; \; + \; \;
{\bf{\Omega}} \; \times \; ({\bf{\Omega}} \; \times \; {\bf{r''}})~~~,
\label{anonimous}
\ee
where the first and the third terms vanish, as the values of relative
deformations $\delta l/l$ are minute in a solid body. The values
of ${\bf v}''\approx \delta l/\tau$ and ${\bf a}'' \approx \delta l/\tau^2$,
where $\tau$ is the period of rotation,
are negligible compared to velocities and accelerations in the inertial
system of reference that by the order of magnitude are, respectively
$l/\tau$ and $l/\tau^2$. The second and the fourth term in 
Eq.~(\ref{anonimous}), referred to the grain frame with unit vectors
(${\bf e}_1$,${\bf e}_2$, ${\bf e}_3$), will read:
\ba
{\bf{\dot{\Omega}}} \; \times \; {\bf{r''}} 
& = &
{{\bf{e}}_1} \; ({\dot{\Omega}}_2 \; z \; \; - 
\; \;  {\dot{\Omega}}_3 \; y) \; \; + \; \; 
{{\bf{e}}_2} \; ({\dot{\Omega}}_3 \; x \; \; -
\; \;  {\dot{\Omega}}_1 \; z) \; \; + \; \;
{{\bf{e}}_3} \; ({\dot{\Omega}}_1 \; y \; \; -
\; \;  {\dot{\Omega}}_2 \; x)  \nonumber  \\
& = &
{{\bf{e}}_1} \; {\omega} \; z \; {\Omega_1} \; \; + \; \; 
{{\bf{e}}_2} \; {\omega} \; z \; {\Omega_2} \; \; + \; \;
{{\bf{e}}_3} \; (\; - \; {\omega} \; y \; {\Omega_2} \; - \;
{\omega} \; x \; {\Omega_1}) \;\;\; ,
\label{26}
\ea
and
\ba
{\bf{\Omega}} \; \times \; ({\bf{\Omega}} \; \times \; {\bf{r''}}) \; \; =\; \;
{\bf{\Omega}} \; ({\bf{\Omega}} \; \cdot \; {\bf{r''}}) \; \; - \; \; 
{\bf{r''}} \; {\bf{\Omega}}^2 \; \; = \nonumber \\
{{\bf{e}}_1} \; \{ 
{\Omega_1} \; ( \; {\Omega_1} \; x \; + \; {\Omega_2} \; y + \; {\Omega_3} \; z
\; ) \; \; - \; \; x \; {\Omega}^2
\} \; \; + \nonumber  \\
{{\bf{e}}_2} \; \{
{\Omega_2} \; ( \; {\Omega_1} \; x \; + \; {\Omega_2} \; y + \; {\Omega_3} \; z
\; ) \; \; - \; \; y \; {\Omega}^2
\} \; \; + \nonumber  \\
{{\bf{e}}_3} \; \{
{\Omega_3} \; ( \; {\Omega_1} \; x \; + \; {\Omega_2} \; y + \; {\Omega_3} \; z
\; ) \; \; - \; \; z \; {\Omega}^2
\}~~~. 
\label{27}
\ea
All in all,
\ba
{\bf{a}} \; \; = \; \; 
{{\bf{e}}_1} \; \{
\; - \; x \; {\Omega_2}^2 \; \; - \; \; x \; {\Omega_3}^2 \; \; + \; \;
y \; {\Omega_2} \; {\Omega_1} \; \; + \; \; z \; {\Omega_3} \; {\Omega_1} 
\; \; + \; \; z \; \omega \; {{\Omega}_{\perp}} \; \cos \; {\omega}t \; 
\}  \nonumber  \\
+ \; \;  
{{\bf{e}}_2} \; \{
\; x \; {\Omega_2} \; {\Omega_1} \; \; - \; \; y \; {\Omega_1}^2 \; \; - \; \;
y \; {\Omega_3}^2 \; \; + \; \; z \; {\Omega_3} \; {\Omega_2}\; \;\
 + \; \; z \; \omega \; {{\Omega}_{\perp}} \; \sin \; {\omega}t \;
\}  \nonumber  \\
+ \; \; 
{{\bf{e}}_3} \; \{
\; x \; {\Omega_1} \; {\Omega_3} \; \; + \; \; y \; {\Omega_3} \; {\Omega_2}  
\; \; - \; \; z \; {\Omega_{\perp}}^2 \; \; - 
\; \; \omega \; {\Omega}_{\perp} 
\; (x \; \cos \; {\omega}t \; + \; y \; \sin \; {\omega}t) \;
\}~~~.
\label{28} 
\ea
We had to present our calculation in great detail because the latter formula 
considerably differs from the one in (Purcell 1979), eq.~28.  
To be doubly sure of our formula being correct, we derived  
it also in the intermediate, ``Eulerian'', system of reference 
(see Appendix~\ref{A}). 

\section{Stresses and Strains Caused by the Precession}

In this section we derive the stresses and strains produced 
in the rotating body by the time-dependent terms in ($\ref{28}$), as only 
these terms influence the rate of energy dissipation in the grain.

With aid of (\ref{rotation}) one can easily split ($\ref{28}$) 
into a time-independent and time-dependent parts:
\be
{\bf{a}} \; \; = \; \; {\bf{a}}_{0} \; \; + \; \; {\bf{a}}_{\mathsf{t}}~~~,
\label{4.1} 
\ee
where
\begin{equation}
{\bf{a}}_{0} \; \; = \; \; - \; ( \; {{\bf{e}}_1} \; x \; \; + \; \; 
                                     {{\bf{e}}_2} \; y \; ) \; 
( \; {\Omega}_3^2 \; \; + \; \; {\frac{1}{2}} \; {\Omega}_{\perp}^2 \; ) \; \;
- \; \; {{\bf{e}}_3} \; z \; {\Omega}_{\perp}^2~~~, 
\label{4.2}
\end{equation}
and
\begin{eqnarray}
{\bf{a}}_{\mathsf{t}} \; \; = \; \; {{\bf{e}}_1} \; \{ \;
{\frac{1}{2}} \; {\Omega}_{\perp}^2 \; x \; \; \cos \; 2{\omega}t \; \; + \; \; 
{\frac{1}{2}} \; {\Omega}_{\perp}^2 \; y \; \; \sin \; 2{\omega}t \; \; + \; \;
z \; {\Omega}_{\perp} \; {\Omega}_3 \; h \; \; \cos \; {\omega}t  
\; \}  \nonumber  \\
+ \; \; 
{{\bf{e}}_2} \; \{
{\frac{1}{2}} \; {\Omega}_{\perp}^2 \; x \; \; \sin \; 2{\omega}t \; \; - \; \;
{\frac{1}{2}} \; {\Omega}_{\perp}^2 \; y \; \; \cos \; 2{\omega}t \; \; + \; \;
z \; {\Omega}_{\perp} \; {\Omega}_3 \; h \; \; \sin \; {\omega}t \; 
\; \}  \nonumber  \\
+ \; \; 
{{\bf{e}}_3} \; \{
{\Omega}_{\perp} \; {\Omega}_3 \; (2 \; - \; h) \; (x \; \; \cos \; {\omega}t 
\; \; + \; \; y \; \; \sin \; {\omega}t \; )
\; \}~~~.
\label{4.3}
\end{eqnarray}
Now we see that dissipation of energy will be taking place at two modes one  
of which will be of double frequency. Hereafter we shall be taking into 
consideration only time-dependent, i.e. produced by 
${\mathbf{a}}_{\mathsf{t}}$, inputs in the stresses and strains.

Following Purcell, we shall model the oblate grain by a prism of sizes
$2a \times 2a \times 2c$, ($c < a$).
The stresses vanishing on the boundaries of our rectangular prism 
read\footnote{Our formulae
for the stresses differ from those in Purcell (1979). 
Purcell's expressions do not include the double-frequency
terms, and they do not vanish on the boundaries of the
grain, which they should do.}:
\be
\sigma_{xx} = \frac{\rho \Omega_\perp^2}{4}  
(x^2 - a^2) \; \cos 2 {\omega} t  \; \; , \; \; \;  
\sigma_{yy} = - \frac{\rho \Omega_\perp^2}{4} (y^2 - a^2) \;  
\cos 2 {\omega} t \; \; , \; \; \;  
\sigma_{zz} = 0~, 
\label{4.5}
\ee
\be
\sigma_{xy} \; \; = \; \; \frac{\rho}{4} \; \Omega_\perp^2 \; 
(x^2 \; \; + \; \; y^2 \; \; - \; \; 2a^2)
\; \; \sin \; 2 {\omega} t \; \; \; ,
\label{4.6}
\ee
\be
\sigma_{xz} \; \; = \; \; \frac{\rho}{2} \; {\Omega_\perp} \; {\Omega_3} \; 
\left[ \; h \; (z^2 \; - \; c^2)
\; \; + \; \; (2 \; - \; h) \; (x^2 \; - \; a^2) \; \right] \; \; 
\cos \; {\omega} t \; \; \; ,
\label{4.7}
\ee
\be
\sigma_{yz} \; \; = \; \; \frac{\rho}{2} \; {\Omega_\perp} \; {\Omega_3} \; 
\left[ \; h \; (z^2 \; - \; c^2)
 \; \; + \; \; (2 \; - \; h) \; (y^2 \; - \; a^2) \; \right] \; \; 
\sin \; {\omega} t \; \; \; ,
\label{4.8}
\ee
which leads to forces per unit volume:
\be
F_x  = {{\partial}_{i}} {{\sigma}_{xi}}  \; \; = 
\frac{\rho}{2} {\Omega_\perp}^2 x \; \cos  2 {\omega} t \; \; + \; \; 
\frac{\rho}{2} {\Omega_\perp}^2 y \; \sin  2 {\omega} t \; \; +  \; \; 
\rho {\Omega_\perp} {\Omega_3} h  z \; \cos  {\omega} t \; \; \; , 
\label{4.9}
\ee
\be
F_y  = {{\partial}_{i}} {{\sigma}_{yi}} \; \; = 
\frac{\rho}{2} {\Omega_\perp}^2 x  \; \sin 2 {\omega} t \; \; - \; \;  
\frac{\rho}{2} {\Omega_\perp}^2 y \; \cos 2 {\omega} t  \; \; + \; \; 
\rho {\Omega_\perp}  {\Omega_3} h z \; \sin  {\omega} t \; \; \; , 
\label{4.10}
\ee
\be
F_z  = {{\partial}_{i}} {{\sigma}_{zi}} = 
\rho {\Omega_\perp} {\Omega_3}  (2  -  h) x  \; \cos \omega t \; \; + \; \; 
\rho {\Omega_\perp} {\Omega_3}  (2  -  h) y  \; \sin \omega t \; \; \; , 
\label{4.11}
\ee
in full accordance with ($\ref{4.3}$). Here $\rho$ is the density of the 
grain material. For a $\; 2a \times 2a \times 2c \;$ prism, the moment of 
inertia $I_3$ and the parameter $h$ are function of the half-sizes $a$ and $c$:
\be
I_3 \;  = \; \frac{16}{3} \; \rho \; a^4 \; c
\label{4.4}
\ee
and 
\be
h \; \; \equiv \; \; \frac{I_3}{I} \; \; = \; \; \frac{2}{1 \; + \; (c/a)^2}  
\; \; \; \; \; , \; \; \; \; \; \; \; \; \; \; \; \; 2 \; - \; h \; \; = 
\; \; \left( \frac{c}{a} \right)^2 \; \frac{2}{1 \; + \; (c/a)^2} \;
\; \; .
\label{4.12}
\ee
According to (3.12), (3.13),
\be
\Omega_3 \; \; = \; \; \Omega_0 \; \cos \theta 
\; \; \; \; \; , \; \; \; \; \; \; \; \; \; \; \; \; 
\Omega_{\perp} \; \; = \; \; \Omega_0 \; h \; \sin \theta \; \; \; ,
\label{4.13}
\ee
$\Omega_0 \; \equiv \; J/I_3$ being the typical angular velocity 
of a grain. Our knowledge of the stresses should be sufficient for computing 
the rate of energy losses in the body. 

Our expressions for the stress-tensor components differ very significantly 
from the appropriate expressions derived in (Purcell 1979). There are two 
reasons for it. The first reason is the afore-mentioned Purcell's 
miscalculation of the acceleration experienced by a point inside the wobbling 
body. (Because of that Purcell completely missed the double-frequency 
contribution to the stress tensor. Later we shall show that this input is of 
the leading order.) The second reason is that Purcell forgot  
to impose proper boundary conditions upon the stress-tensor components: his 
stresses fail to vanish on the boundaries of the body.

Our goal now is to calculate the average deformation-caused energy, stored in 
the tumbling body, and to estimate the energy dissipation, using an 
appropriate quality factor. We shall take into account that the deformation of
the grain is neither purely elastic nor purely plastic, but is a superposition
of the former and the latter. It is then to be described by the tensor 
$\epsilon_{\it{ij}}$ of {\textit {viscoelastic}} strains and by the velocity 
tensor consisting of the time-derivatives $\dot{\epsilon}_{\it{ij}}$. The 
stress tensor will now be separated into two components: the elastic stress 
and the plastic (viscous) stress:
\be
\sigma_{\it{ij}} \; \; = \; \; \sigma_{\it{ij}}^{(e)} \; \; + 
\sigma_{\it{ij}}^{(p)}~~~,
\label{4.18}
\ee
where the components of the elastic stress tensor are interconnected with 
those of the strain tensor (Landau and Lifshitz 1976):
\be
\epsilon_{ij} \; \; = \; \; \delta_{ij} \; \; \frac{Tr \; 
\sigma^{(e)}}{9 \; K} \; \; + \; \; 
\left( \; \sigma_{\it{ij}}^{(e)} \; \; - \; \; \frac{1}{3} \; \; \delta_{ij} 
\; \; Tr \; \sigma^{(e)} \right) \; \frac{1}{2 \; \mu} \; \; \; ~~~,
\label{4.19}
\ee
\be
\sigma_{ij}^{(e)} \; \; = \; \; K \; \delta_{ij} \; \; Tr \; \epsilon \; \; + 
\; \; 2 \; \mu \; \left(\epsilon_{ij} \; - \; \frac{1}{3} \; \delta_{ij} \; 
\; Tr \; \epsilon \right)~~~,
\label{4.20}
\ee
$\mu$ and $K$ being the {\textit{isothermal}} shear and bulk moduli, and $Tr$ 
standing for the trace of a tensor. Components of the plastic stress are 
connected with the strain derivatives as 
\be
\dot{\epsilon}_{ij} \; \; = \; \; \delta_{ij} \; 
\frac{Tr \; \sigma^{(p)}}{9 \zeta} \; \; + \; \; \left(\sigma_{ij}^{(p)} \; - 
\; \frac{1}{3} \; \delta_{ij} \;\; Tr \; \sigma^{(p)} \right) \; 
\frac{1}{2 \eta}~~~,
\label{4.21}
\ee
\be
\sigma_{\it{ij}}^{(p)} \; \; = \; \; \zeta \; \delta_{ij} \; \; 
Tr \; {\dot{\epsilon}} \; \; + \; \; 2 \; \eta \; \left(\dot{\epsilon}_{ij}
\; - \; \frac{1}{3} \; \delta_{ij} \; \; Tr \; \dot{\epsilon} \right)~~~, 
\label{4.22}
\ee
where  $\; \eta \;$ and $\; \zeta \;$ are the shear and stretch viscosities.

\section{Dynamics of a tumbling body}

The kinetic energy of a rotating body reads, according to (\ref{2.5}), 
(\ref{2.6}), (\ref{rotation}), and (\ref{om4}):
\begin{eqnarray}
E_{rot} \; \; = \; \; \frac{1}{2} \; (I_1 \; \Omega_1^2 \; + \; I_2 \; \Omega_2^2 \; + \; I_3 \; \Omega_3^2) \; \; 
= \; \; \frac{1}{2} \;
[I \; \Omega_{\perp}^2 \; + \; I_3 \; {\Omega_3}^2] \; \; = 
\nonumber\\
= \; \; \frac{1}{2} \; \; \left[ \; \frac{1}{I} \; \; \sin^2 \; \theta \; \; + \; \; 
\frac{1}{I_3} \; \; \cos^2 \; \theta \; \right] \; J^2
\label{5.1}
\end{eqnarray}
wherefrom
\be
\frac{dE_{rot}}{d\theta} \; \; = \; \; 
\frac{J^2}{I_3} \; (h \; - \; 1) \; \; \sin \; \theta \; \; \cos \; \theta 
\; \; =  \; \; \omega \; J \; \; \sin \; \theta \; \; \; .
\label{5.2}
\ee
Formula (\ref{5.2}) provides an insight in how the change of rotational energy 
yields a change of $ \; \theta$. If we calculate the rate of 
energy losses, $\dot{E}_{rot}$, it will be easy to find the
rate of alignment       
$\; \dot{\theta} \;$ as $\; ({dE_{rot}}/{d\theta})^{-1}\dot{E}_{rot} \;$.
How to compute $\dot{E}_{rot}$?
The rotational energy changes via the inelastic dissipation, so that 
\be
\dot{E}_{rot} \; = \; \dot{W} 
\label{5.3}
\ee
Hence what we have to find is the rate of the elastic-energy losses $\;
\dot{W}$. Then, with aid of (\ref{5.2}) we shall calculate the rate of 
alignment:
\be
\frac{d\theta}{dt} \; = \; \left(\frac{dE_{rot}}{d\theta}\right)^{-1} 
\frac{dE_{rot}}{dt} \; = \; \left( \omega \; J \; \; \sin \; \theta 
\right)^{-1} \; \dot{W}
\label{5.4}
\ee
In our case, dissipation is taking place at two modes:
\be
\dot{W} \; = \; \dot{W^{({\omega})}} \; + \; \dot{W^{(2{\omega})}} \; = \; 
\omega \; \frac{W^{({\omega})}}{Q^{({\omega})}} \; + \; 2 \; \omega \; 
\frac{W^{({2\omega})}}{Q^{({2\omega})}} \; \approx \; \frac{\omega}{Q} \; 
\left\{ W^{({\omega})} \; + \; 2 W^{({2\omega})} \protect\right\}
\label{5.5}
\ee
where we used the fact that the 
quality factor is almost frequency-independent: $\; Q^{({2\omega})} \, 
\approx \, Q^{({\omega})} \, \approx \, Q \;$.

\section{Rate of Energy Dissipation at low temperatures}

Before pursuing to our calculations of the energy-dissipation rate $\; 
\dot{W} \;$, several prefatory notes will be in order. As well known, at low 
temperatures materials are fragile: when the deformations exceed some critical
threshold, the body will rather break than flow. At the same time, at these 
temperatures the materials are elastic, provided the deformations are 
{\textit{beneath}} the said threshold: the sound absorption, for example, is 
almost exclusively due to the thermal conductivity rather than to the 
viscosity. These facts may be summarized like this: at low temperatures, the 
viscosity coefficient $\; \eta \;$ has, effectively, two values: one value - 
for small deformations (and this value is almost exactly zero); another value 
- for larger-than-threshold deformations (and that value is  
high\footnote{Effectively it may be put infinity because, as 
explained above, the body will rather crack than demonstrate fluidity.}).

At high temperatures materials become plastic, which means that the shear 
viscosity $\; \eta \;$ gets its single value, deformation-independent in the 
first approximation. On the one hand, this value will be far from zero (so 
that the scattering of vibrations will now be predominantly due to the 
viscosity terms, not due to the thermal conductivity). On the other hand, this
value will not be that high: a plastic body will rather yield than break. 
All this is certainly valid for the stretch viscosity $\; \zeta \;$ as well.

Since at low temperatures the bodies manifest, for small displacements, no 
viscosity ($\; \omega \eta \, \sim \, \omega \zeta \; \ll \;  \mu \, \sim \, K
\; $), the stress tensor will be approximated, to a high accuracy, by its 
elastic part: instead of the system (\ref{4.18}) - (\ref{4.22}) we shall 
simply write:
\be
\epsilon_{ij} \; \; = \; \; \delta_{ij} \; \; \frac{Tr \; 
\sigma}{9 \; K} \; \; + \; \; 
\left( \; \sigma_{\it{ij}} \; \; - \; \; \frac{1}{3} \; \; \delta_{ij} 
\; \; Tr \; \sigma \right) \; \frac{1}{2 \; \mu} \; \; \; ~~~,
\label{6.1}
\ee
This will enable us to derive an expression for the elastic energy stored in 
a unit volume of the precessing body:
\ba
dW/dV \; = \; \frac{1}{2} \; \epsilon_{ij} \; \sigma_{ij} \; =\;\frac{1}{4} \; 
\left\{ \left(\frac{2 \; \mu}{9\; K} \; - \; \frac{1}{3} \protect\right) \, 
\left(Tr \; \sigma \protect\right)^2 \; + \; \sigma_{ij} \,  \sigma_{ij} 
\protect\right\} \; \approx 
\nonumber \\
\nonumber \\
\approx \; \; \frac{1}{4\mu}\;\left\{\; - \; \frac{1}{5} \, \left(Tr \; \sigma 
\protect\right)^2 \; + \; \sigma_{xx}^2\; + \; \sigma_{yy}^2 \; + \; 
\sigma_{zz}^2 \; + \; 2 \, \left( \sigma_{xy}^2 \; + \; \sigma_{yz}^2 \; + \; 
\sigma_{zx}^2 \protect\right) \protect\right\}
\label{6.2}
\ea 
where we made use of the expressions connecting the shear and bulk moduli with
the Young modulus $E$ and Poisson's ratio $\sigma$: since $\; K \, = \, E/[3(1
-2\sigma] \;$ and $\; \mu \, = \, E/[2(1+\sigma] \;$ then $\; 2 \mu/(9K) \, - 
1/3 \, = \, - \, \sigma /(1+\sigma ) \;$. As for frozen solids Poisson's 
ratio $\sigma$ is typically about $0.25$, we put  $\; 2 \mu/(9K) \, - 1/3 \; 
\approx \; - \, 1/5 \;$. Mind that the body is trembling at two frequencies: 
$\; \omega \;$ and $\; 2 \, \omega \;$. Anticipating the different rates of 
dissipation at these two modes, we shall split the total elastic energy into 
two parts: 
\be
dW/dV \; = \; dW^{(\omega)}/dV \; + \; dW^{(2\omega)}/dV
\label{6.3}
\ee
where, according to (\ref{4.5}) - (\ref{4.8}), 
\be
dW^{(\omega)}/dV \; = \; \frac{1}{2\mu}\;\left\{ \sigma_{yz}^2 \; + \; 
\sigma_{zx}^2 \protect\right\} 
\label{6.4}
\ee
and 
\be
dW^{(2\omega)}/dV \; = \; \frac{1}{4\mu} \; \left\{\; - \; \frac{1}{5} \,
\left(Tr \; \sigma 
\protect\right)^2 \; + \; \sigma_{xx}^2\; + \; \sigma_{yy}^2 \; + \; 
2 \sigma_{xy}^2  \protect\right\} \; \; \; .
\label{6.5}
\ee
>>From now on we shall be interested in the energies averaged over several 
periods of the precession. Therefore we shall substitute $\; \sin^2... \;$ and
$\; \cos^2... \;$ by $1/2$, and shall omit expressions $\; \sin... \, \cos...\;
$, $\; \sin \omega t \, \sin 2 \omega t \;$ and $\; \cos \omega t \, \cos 2 
\omega t \;$. With the above reservation being beared in mind, 
expressions (\ref{4.5}) - (\ref{4.8}) for the stresses will yield:
\be
\left(Tr \; \sigma \protect\right)^2 \; = \; \left(\frac{\rho \; 
\Omega^2_{\perp}}{4}\protect\right)^2 \; a^4 \; \left(x_1^2 \; - \; 
y_1^2\protect\right)^2 \; \frac{1}{2} \; \; \; ,
\label{6.6}
\ee
\be
\sigma^2_{xx} \; = \; 
\left(\frac{\rho \; \Omega^2_{\perp}}{4}\protect\right)^2 \; a^4 \; 
\left(x_1^2 \; - \; 1\protect\right)^2 \; \frac{1}{2} \; \; \; ,
\label{6.7}
\ee
\be
\sigma^2_{yy} \; = \; 
\left(\frac{\rho \; \Omega^2_{\perp}}{4}\protect\right)^2 \; a^4 \; 
\left(y_1^2 \; - \; 1\protect\right)^2 \; \frac{1}{2} \; \; \; ,
\label{6.8}
\ee
\be
\sigma^2_{xy} \; = \; 
\left(\frac{\rho \; \Omega^2_{\perp}}{4}\protect\right)^2 \; a^4 \; 
\left(x_1^2 \; + \; y^2_1 \; - \; 2\protect\right)^2 \; \frac{1}{2} \; \; \; ,
\label{6.9}
\ee
\be
\sigma^2_{xz} \; = \; 2 \; \gamma \; 
\left(\frac{\rho \; \Omega^2_{\perp}}{4}\protect\right)^2 \; a^4 \; 
\left(x_1^2 \; + \; z^2_1 \; - \; 2\protect\right)^2 \; \; \; ,
\label{6.10}
\ee
\be
\sigma^2_{yz} \; = \;  2 \; \gamma \; 
\left(\frac{\rho \; \Omega^2_{\perp}}{4}\protect\right)^2 \; a^4 \; 
\left(y_1^2 \; + \; z^2_1 \; - \; 2\protect\right)^2 \; \; \; ,
\label{6.11}
\ee
where 
\be
x_1 \; \; \equiv \; \; x/a \; \; \; , \; \; \; \; \; \; y_1 \; \; \equiv \;
y/a \; \; \; \; \; , \; \; \; \; z_1 \; \; \equiv \; \; z/a \; \; \; ,
\label{6.12}
\ee
and
\be
\gamma \; \; \equiv \; \; \left( \frac{\Omega_{3}}{\Omega_{\perp}}
\protect\right)^2 \; (2 \; - \; h)^2 \; \; = \; \; 
\left( \frac{\Omega_{3}}{\Omega_{\perp}}
\protect\right)^2 \; h^2 \; \left(\frac{c}{a} \protect\right)^4 \; \; = 
\; \; \frac{(c/a)^4}{\tan^2 \theta} \; \; \; .
\label{6.13}
\ee
According to (\ref{5.5}), what we shall need is the sum $\; W^{({\omega})} \; +
\; 2 \, W^{({2\omega})}\;$. To get it, we shall integrate over the volume: 
\ba
W^{({\omega})} \; + \; 2 W^{({2\omega})} \; = \; \int^{a}_{-a} \, dx \; 
\int^{a}_{-a} \, dy \; \int^{c}_{-c} \, dz \; \left\{ dW^{({\omega})}/dV \; + 
\; 2 \; dW^{({2\omega})}/dV  \protect\right\} \; =
\nonumber \\
= \; a^2 \; c \; \int^{1}_{-1} \, dx_1 \; \int^{1}_{-1} \, dy_1 \; 
\int^{1}_{-1} \, dz_1 \; \frac{1}{2\mu}\left\{ 
\; - \; \frac{1}{5} \,
\left(Tr \; \sigma \protect\right)^2 \; + \; \sigma_{xx}^2\; + \; 
\sigma_{yy}^2 \; + \; 2 \sigma_{xy}^2 \; + \; \sigma_{yz}^2 \; + \; 
\sigma_{zx}^2 \protect\right\} \; \; \; . 
\label{6.14}
\ea
Plugging of (\ref{6.6}) - (\ref{6.11}) in the above expression entails:
\ba
W^{({\omega})}\;+\;2 W^{({2\omega})} \; =\;2^{-5}\;\; \frac{a^6 \; c}{\mu} \; 
\left(\rho \; \Omega^2_{\perp}\protect\right)^2 \; 
\int^{1}_{-1} \, dx_1 \; \int^{1}_{-1} \, dy_1 \; 
\int^{1}_{-1} \, dz_1 \; \left\{ \right.
\; - \; \frac{1}{10} \; \left(x_1^2 \; - \; y_1^2 \protect\right)^2 \; + 
\nonumber \\
+ \; \frac{1}{2} \; \left(x_1^2 \; - \; 1 \protect\right)^2 \; + 
\; \frac{1}{2} \; \left(y_1^2 \; - \; 1 \protect\right)^2 \; +  
\; \left(x_1^2 \; + \; y_1^2 \; - \; 2 \protect\right)^2 \; +
\nonumber \\
+ \; 2 \; \gamma \; \left(x_1^2 \; + \; z_1^2 \; - \; 2 \protect\right)^2 \; +
\; 2 \; \gamma \; \left(y_1^2 \; + \; z_1^2 \; - \; 2 \protect\right)^2
\left. \protect\right\} \; \approx
\nonumber \\
\approx \; \frac{a^6 \; c}{\mu} \; \left(\rho \; 
\Omega^2_{\perp}\protect\right)^2 \; 2^{-5} \; (63 \; \gamma \; + \; 20)
\label{6.15}
\ea
where the numerics was performed by means of the Maple software.
The product $\;\rho^2\Omega_{\perp}^4\;$ emerging in the above expression may 
be cast in a form wherein its temperature-dependence 
becomes manifest. 
According to  ($\ref{4.4}$), ($\ref{4.12}$), and ($\ref{4.13}$), 
\begin{eqnarray}
\rho^2 \; \Omega_{\perp}^4 \; \; = 
\frac{a^{-8} \; c^{-2}}{ \left[1+(c/a)^2  \protect\right]^4} \; 
\frac{9}{4} \; \left( \beta \; kT_{\rm gas} \protect\right)^2 \; \sin^4
\theta \; \; \; \; \; \; \; 
\label{6.16}
\end{eqnarray}
where $k$ signifies the Boltzmann constant and $\beta$ stands for the
parameter of suprathermality: 
\be
\frac{I_3 \; \Omega_0^2}{2} \; \; = \; \; \frac{J^2}{2I_3} \; \; = \; \; 
\beta \; kT_{\rm gas}
\label{6.17}
\ee
The cosmic dust is thermal when $\beta = 1$. This way of writing the
energies enables one to express these in terms of the temperature of
the surrounding gas, $T_{\rm gas}$:
\be
W^{({\omega})} \; + \; 2 \; W^{({2\omega})} \; = \; \frac{3^2 \; 2^{-7} \; 
(63 \; \gamma \; + \; 20)}{\mu} \;
\frac{a^{-2} \; c^{-1}}{ \left[1+(c/a)^2  \protect\right]^4} \; 
\left( \beta \; kT_{\rm gas} \protect\right)^2 \; \sin^4 \theta
\label{6.18}
\ee
Together, (\ref{5.4}), (\ref{5.5}) and (\ref{6.18}) give:
\be
d \theta/dt \; = \; - \; \frac{3^2 \; 2^{-7} \; 
(63 \; \gamma \; + \; 20)}{\mu \; Q \; J} \;
\frac{a^{-2} \; c^{-1}}{ \left[1+(c/a)^2  \protect\right]^4} \; 
\left( \beta \; kT_{\rm gas} \protect\right)^2 \; \sin^3 \theta
\label{6.19}
\ee
According to (\ref{6.17}) and (\ref{4.4}),
\be
J \; = \; \left(2 \; \beta \; k \; T_{gas} \; I_3 \protect\right)^{1/2} \;
= \; \left( \frac{32}{3} \; \beta \; k \; T_{gas} \; \rho \; a^4 \; c 
\protect\right)^{1/2} \; = \; 2^{5/2} \; 3^{-1/2} \; a^{2} \; c^{1/2} \; 
\rho^{1/2} \; \left( \beta \; k \; T_{gas} \protect\right)^{1/2} 
\label{6.20}
\ee
Substitution of the latter in the former will give us the final expression for 
the alignment rate:
\be
d \theta/dt \; = \; - \; \frac{a^{-5.5}}{[1+(c/a)^2]^4} \; 
\left(\frac{a}{c}\protect\right)^{3/2} \; 
\left( \beta \; kT_{\rm gas} \protect\right)^{3/2} \; 
\frac{\sin^3 \theta}{\mu \; Q \; \rho^{1/2}} \;
 3^{2.5} \; 2^{-9.5} \; \left(63 \; (c/a)^4 \; \cot^2 \theta \; + \; 20
\protect\right) 
\label{6.21}
\ee
Simply from looking at this formula one can conclude that the 
major-inertia axis slows down its alignment for $\; \theta \; $ approaching 
zero. This feature looks physically reasonable. 

On physical grounds, one may also expect that the alignment rate vanishes 
for $\; h \;$ approaching unity (i.e., when the body lacks an axis 
of maximal inertia). Besides, one may expect the major axis of the grain 
to linger in the position $\; \theta \; \approx \; \pi /2 \;$, 
as if ``hesitating'' to whether to start aligning along or opposite its 
angular momentum. However, none of the latter two features seems to be 
instilled 
into (\ref{6.21}): it may seem from this formula that $\; d \theta /dt \;$ is 
$\; h-$independent, and that the major axis leaves the initial position $\; 
\theta = 
\pi/2 \;$ with a finite angular velocity. To dispel these illusions, simply 
recall that our treatment is valid only for as long as the rotation and 
precession are fast motions compared to the alignment: $\; {\dot{\theta}} \ll
\omega \;$. 

All these subtleties are anyway irrelevant when one merely wants to 
estimate the relaxation time, i.e., the time required for the maximal-inertia 
axis of an oblate cosmic-dust grain to be considerably shifted toward 
alignment with the angular momentum:
\begin{equation}
t_{i} \; \; \simeq \; \; 
\left( \; \left\langle \frac{d\theta}{dt} \right\rangle \; \right)^{-1} 
\; \; \simeq \; \; - \int_{\pi/2}^{\delta} \; 
\frac{d \theta}{d \theta/dt} \; \; \; .
\label{6.22}
\end{equation}
where $\delta$ is introduced to avoid the divergence associated with the
``slow finish''. (One can take, for example, $\; \delta \; = \; \pi/8 \;$.) A 
particular choice of $\delta$ will bring into the expression for 
$t_{\mathsf{i}}$ some numerical factor of order unity. Since we want nothing 
more but a rough estimate for $t_{\mathsf{i}}$, we shall approximate 
$t_{\mathsf{i}}$ simply by the inverse $\; d \theta /dt \;$ evaluated in the 
middle of the interval, at $\theta \; = \; \pi/4$. Then 
\begin{eqnarray}
t_{i} \; 
&\approx & a^{5.5} \; [1+(c/a)^2]^4 \; 
\left(\frac{c}{a}\protect\right)^{3/2} \; 
\left( \beta \; kT_{\rm gas} \protect\right)^{-3/2} \; 
\mu \; Q \; \rho^{1/2} \; \frac{2^{11} \; 3^{-2.5}}{63(c/a)^4+20}
\label{6.23}
\end{eqnarray}

The quantity $\; (c/a)^{3/2} \; [1+(c/a)^2]^4 \; [63 \, (c/a)^4 \, + \, 
20]^{-1} \;$ is a steep function of parameter $\; (c/a) \;$. For $\; c/a = 1/2
\;$ it equals to $\; 4.3 \, \times \, 10^{-2} \;$, while for $\; c/a = 1/10 \;$ it will be  $\; 1.6 \, \times \, 10^{-3} \;$.   

The above formula for the typical time of alignment is the main 
result of our article. Now we must think of the possible values 
for the material parameters involved. The values may depend both on the 
temperature $\; T_{\rm grain} \;$ of the cosmic-dust grains, and on the frequency 
of the precession. To start with, temperature- and frequency-caused variations
of the density $\; \rho \;$ may be neglected, as they are to be small anyway. 
This way, we can use the (static) densities appropriate to the room 
temperature and pressure. 

Now, consider the isothermal shear modulus $\; \mu \;$. The tables of 
physical quantities would provide its values for the room temperature and 
atmospheric pressure, and for quasistatic regimes solely. 
As for the possible frequency-related effects in materials (the so-called 
ultrasonic attenuation), these become noticeable only at  
frequencies higher than $\; 10^8 \; Hz \;$ (see 
section 17.7 in Nowick and Berry 1972). Another 
fortunate circumstance is that the pressure-dependence of 
the elastic moduli is known to be weak (Ahrens 1995). Besides, the elastic 
moduli of solids are known to be insensitive to temperature variations, as 
long as these variations are far enough from the melting point. The value of 
$\; \mu \;$ may increase by several percent when the temperature is 
drops from the room temperature to 10 K.  Dislocations don't affect the
elastic moduli either. Solute elements have very little effect on 
moduli in quantities up to a few percent\footnote{Beyond that, one might 
assume that the moduli vary linearly with substitutional impurities (in 
which the atoms of the impurity replace those of the hosts). However hydrogen 
is not like that: it enters the interstices between the atoms of the host, 
and has marginal effect on modulus.}. As for the role of the possible 
porosity, the elastic moduli scale  as the square of the relative density. 
For porosities up to about $20 \; \% ~\,$, this is not of much relevance 
for our estimates\footnote{We are deeply thankful to Michael Ashby and Michael
Aziz, who consulted us on all these subtle topics.}. 

According to (Ryan and Blevins 1987), 
at $\; T \; \simeq \; 20 \, K \;$, the share modulus value for silicate 
$\mu \; \approx \; 10^{10} \; Pa \; \; $ 
and density is $\rho^{\rm silicate} \approx \; 2500 \; kg \; 
m^{-3} $. 
The temperature of subthermal rotating that according
to Lazarian and Draine (1997) is important during crossovers
is $T_{\rm rot} \; \simeq \; 20 \, K \,$. 

Now, several words on the choice of values of the $\; Q-$factor will be in 
order. Purcell refers to (Krause 1973) wherein a review of acoustic 
dissipation in silicates is presented. According to (Krause 1973), for 
frequencies varying from $\; 50 \; kHz \;$ to $\; 27 \; GHz \;$, and 
temperatures varying from $\; 10 \; K \;$ to $\; 50 \; K \;$, the measured 
values of $\; Q \;$ range between $\; 400 \;$ and $\; 2000 \;$. 
 We would tend to believe that interstellar grains
are almost certainly have plenty of cracks and 
defects. As known from seismology, for real silicate rocks the $\; 
Q-$factor is typically between $\; 150 \;$ and $\; 300 \;$. 
Therefore we believe  that
$Q^{({\rm sil})} \; < \; 400$.
We are not familiar with the measurements of $Q$ for carbonatious materials,
but it is likely that $Q$ for them will be lower than for silicates.

Therefore, for the silicate grains with axis ratio 1:2 we get:
\be
t_{i} \; \approx \;  8 \; \times \; 10^9 \; \frac{1}{\beta^{3/2}} \; \left( \frac{a}{10^{-7} \; m}\protect\right)^{5.5} \; \;,
\label{6.26}
\ee
while for graphite grain we assume that the axis ratio is 1:10 $Q=100$
and obtain
\be
t_{i} \; \approx \; 7 \; \times \; 10^6 \; \frac{1}{\beta^{3/2}} \; 
\left( \frac{a}{10^{-7} \; m}\protect\right)^{5.5} \; \;,
\label{6.27}
\ee
where the values of $\; a \;$ are supposed to be in $metres$ while $t$ is in 
$seconds\;$ (so that for example for $\; a \, = \, 10^{-7} \, m \;$, $\, c/a 
\, = \, 1/2 \,$ and $\, \beta \, = \, 1 \,$ the typical time will be  $\,
8 \, \times \, 10^{9} \, s\,$ ) .

We have not attempted to calculate the dissipation rates for grains formed by 
loose aggregates of smaller particles.
The inelastic relaxation within such grains may be orders of magnitude
more efficient due to friction between parts of it (``effective viscosity'').
It can well dominate the Barnett relaxation even for much larger grains.

\protect\section{Dynamics of Prolate Cosmic-Dust Grains: Libration.}

At firts glance, the dynamics of a freely-spinning prolate body obeys the same 
principles as the dynamics of an oblate one: the axis of maximal inertia will 
tend to align itself parallel to the angular momentum. If we model a prolate 
body with a symmetric top, it will be once again convenient to choose it be a 
prism of dimensions $2a \times 2a \times 2c$, though this time half-size $\, c
 \,$ are larger than $\, a \,$, and therefore $\; I_3 \; = \; I_2 \; > \; I_1 
\;$. Then all our calculations {\textit{formally}} remain in force, 
up to formula (\ref{5.2}): since now the factor $h-1=[1-(c/a)^2]/[1+(c/a)^2]$ 
becomes negative, the right-hand side in (\ref{5.2}) will change its sign:
\be
\frac{dE_{rot}}{d\theta} \; \; = \; \; 
\frac{J^2}{I_3} \; (h \; - \; 1) \; \; \sin \; \theta \; \; \cos \; \theta 
\; \; =  \; - \; \omega \; J \; \; \sin \; \theta \; \; \; .
\label{7.1}
\ee
Thereby formula (\ref{5.4}) will also acquire a ``minus'' sign in its 
right-hand side:
\be
\frac{d\theta}{dt} \; = \; \left(\frac{dE_{rot}}{d\theta}\right)^{-1} 
\frac{dE_{rot}}{dt} \; = \; - \; \left( \omega \; J \; \; \sin \; \theta 
\right)^{-1} \; \dot{W}
\label{7.2}
\ee
Formulae (\ref{5.3}) and (\ref{5.5}) will remain unaltered. 
Eventually, by using (\ref{7.2}) and (\ref{5.5}), we shall arrive to a formula 
that differs from (\ref{6.21}) only by a sign, provided we keep the notation 
$\theta$ for the angle between $\mathbf{J}$ and the body-frame axis 3 
(parallel to dimension $2c$):
\be
d \theta/dt \; = \; - \; \frac{a^{-5.5}}{[1+(c/a)^2]^4} \; 
\left(\frac{a}{c}\protect\right)^{3/2} \; 
\left( \beta \; kT_{\rm gas} \protect\right)^{3/2} \; 
\frac{\sin^3 \theta}{\mu \; Q \; \rho^{1/2}} \;
 3^{2.5} \; 2^{-9.5} \; \left(63 \; (c/a)^4 \; \cot^2 \theta \; + \; 20
\protect\right) 
\label{7.3}
\ee
This looks like axis 3 tends to stand orthogonal to 
$\mathbf{J}$, which seems to be so natural since axis 3 is now not the 
maximal-inertia but the minimal-inertia axis. 

Alas, all this nice extrapolation of the oblate-body-applicable
approach to a prolate-body case is of no practical interest, bacause
in reality an infinitesimally small deviation between the values of $\; I_2 
\;$ and $\; I_3 \;$ leads to a considerably different type of wobble:
the so-called libration (Synge \& Griffith 1959). This phenomenon will
be comprehensively discussed in our next article. 

\protect\section{Discussion}

\subsection{Comparison with the Barnett relaxation}

Another important mechanism of internal relaxation, i.e. the Barnett
relaxation, dissipates the energy via oscillating magnetization that
arising from angular velocity precession in grain body coordinates.
Lazarian \& Draine (1997) provide an estimate 
$t_B\approx A/\omega^2$, where 
$A\approx 7.1 \times 10^{17}(a/10^{-5}~{\rm cm})^2
$~s$^{-1}$ for an oblate grain with 2:1 axis ratio. 
For such grains the Barnett dissipation dominates
for grains with $a<2\times 10^{-6}$~m. However, this is not true
for a larger axis ratio. Indeed, according to Lazarian \& Draine (1997)
for $a/c\gg 1$ $t_{\rm B}$ scales as $(a/c)^6$, while we found
above that $t_i$ scales as $(c/a)^{3/2}$. Therefore fore grains
with 4:1 axis ratio the inelastic relaxation dominates if $a>10^{-7}$~m.
 For grains less than $3\times
10^{-8}$~m the
inelastic relaxation becomes more important. 
It also dominates for grains constituted by loose aggregates
of smaller particles. The exact value of the $Q$-factor and therefore of
relaxation rates would depend
on the structure and properties of the aggregate.

For grains with axis ratio 2:1 and radii $10^{-7}$~m $< a< 2\times 10^{-6}$~m
the Barnett relaxation
is dominant when grains rotate thermally. Suprathermally rotating grains
stochastically undergo spin-ups and spin-downs and during crossovers
rotate with subthermal velocities.
Although short, in terms of grain alignment, 
crossovers, are the most important moments of grain dynamics. Grains
are marginally susceptible to the randomization via gaseous bombardment
when they rotate suprathermally (Purcell 1979). Our calculations of
the inelastic relaxation efficiency show that the Barnett relaxation
is the dominant process that determines internal dissipation during
crossovers within grains with $a>10^{-7}$~m. Therefore results
on paramagnetic and mechanical alignment obtained for such grains
in Lazarian \& Draine (1997) stay unaltered provided that $Q$ factors
of interstellar grains are as high as they were chosen in this paper.

In this paper we assumed that grains are axially symmetric (and
oblate). For grains
of arbitrary shape accelerations will be higher as it tumbles. Therefore
the inelastic relaxation is bound to increase and a relaxation time to
decrease by a numerical factor that will depend on the grain shape.
This subject is beyond the scope of the present paper.

Our treatment of the inelastic relaxation ignored thermal fluctuations in
grain material. In reality, for finite grain temperatures, $\theta$
fluctuates and the thermal distribution proportional to the
Boltzmann factor $\exp(-E_{kin}(\theta)/kT)$ is established as
$t\rightarrow \infty$. To describe the transient processes of alignment
one can solve Fokker-Planck equation as it is done in Lazarian
\& Roberge (1997) in the case of the Barnett relaxation, but to use 
coefficients derived in Appendix $E$.

\subsection{Alignment and Internal dissipation}

Several mechanisms of cosmic-dust alignment are known. They constitute 
three major types: mechanical, paramagnetic and via radiative torques.
All of them appeal to internal relaxation that enables the alignment
of the angular momentum with the axis of the maximal moment of inertia, 
henceforth the axis of major inertia. The degree of achievable alignment
depends on whether grains rotate thermally
\footnote{In Landau and Lifshitz (1969, section 26) it was shown that the 
equilibrium position of an inelastic body corresponds to its 
maximal-inertia axis being parallel to the angular momentum. However in 
reality this statement remains true only up to thermal fluctuations: the 
$\it rms$ of the angular deviation is $\Delta \theta = 
\sqrt{\it{I}kT/\bf{J}^{2}}$, $\bf J$ and $\it {I}$ being the angular 
momentum and the moment of inertia (Lazarian 1994, Lazarian and Roberge 
1997).}  (the average rotational energy of a grain is of the order of 
the kinetic energy of the surrounding gas)  or 
suprathermally (spinning with energies much exceeding $kT$).

We note that all the types of mechanisms below provide the alignment of 
the grain angular momentum in respect to the magnetic field.
This happens because grains swiftly precess about magnetic field lines. 
This precession is 
called into being by the interaction of the grain's magnetic moment with the 
field. The said magnetic moment is generated by the Barnett effect
and is thereby parallel to the angular velocity. For a 
typical cosmic-dust grain ($\sim~10^{-5}$~cm) in a typical interstellar 
magnetic field (5~$\mu$G) the period of Larmor precession is less 
than a week (Purcell 1979), 
which is much less than the typical time of alignment.

To relate the alignment
of angular momentum to the polarization produced by grains, one has 
to know the alignment of grain axis, which is determined 
by the internal dissipation. 
 However, the alignment of angular momentum, in its turn also
depends on the alignment of grain axis. In the cases of
thermally rotating grains, e.g. Davis \& Greenstein (1951) and Gold (1952)
alignment mechanisms  it is sufficient to know that 
the rate of internal dissipation exceeds 
the rate of alignment. In cases of suprathermally rotating grains
more precise estimates
of the internal relaxation time are necessary. The latter include
paramagnetic alignment of suprathermally rotating grains (Purcell 1979,
Spitzer \& McGlynn 1979, Lazarian \& Draine 1979),
crossover and cross sectional mechanical alignment (Lazarian 1995,
Lazarian \& Efroimsky 1996, Lazarian, Efroimsky \& Ozik 1996)
and the alignment via radiative torques (Draine \& Weingartner 1996, 1997).
Our finding that for some grains inelastic relaxation dominates is important
and will be accounted in the quantitative studies of alignment elsewhere.

\subsection{Comparison with the earlier work}

Our treatment differs from that in (Purcell 1979) in a number of points. 

I. We have obtained different expressions for stresses. In particular, our 
stresses do obey the boundary conditions: they vanish on the grain boundary 
(while the stresses in (Purcell 1979) fail to do so).

II. We have found energy dissipation at the double frequency. It is possible 
to show that the double-frequency dissipation provides a leading contribution
\footnote{For example, in formula (\ref{7.3}) the term $\; 63(c/a)^4 \, 
\cot^2 \theta \;$ in the sum in brackets is due to damping at the precession 
frequency $\; \omega \;$, while the term $\; 20 \;$ is due to damping at the 
double frequency. Evidently, even for $\; c/a \approx 1/2 \;$ the double 
frequency provides the leading input, while for $\; c/a \approx 1/10 \;$ 
its input becomes absolutely overwhelming}. The dissipation at the double 
frequency was not considered in (Purcell 1979).

As a result, our final expression for the same $\; Q$-factor predicts higher 
inelastic relaxation efficiency than the corresponding one in (Purcell 1979). 
To make a comparison
we rewrite our formulae in a form similar to that in Purcell (1979).
Then our expression (\ref{6.21}) would read:
\be
\left( \frac{d \theta}{dt} \protect\right)_{_{\mathit{our \; \; result}}} \; 
\approx \; - \; \frac{8}{5} \; \frac{\Omega^3_0 \; \rho \;
a^2}{ \mu \; Q \; \left[1 \; + \; (c/a)^2 \protect\right]^4} \; 
\left(0.4 \; (c/a)^4 \; \cos^2 \theta \; \sin \theta \; +\;0.1 \protect\right)
\label{8.1}
\ee
In the appropriate expression presented in Purcell's article, the second term 
in brackets was missing (because Purcell missed the contribution from the 
dissipation at the double-frequency mode), while the first term is about 2.5 
times larger (because Purcell's calculation of the stress tensor ignored the 
boundary conditions upon stresses).
%\be
%\left( \frac{d \theta}{dt} \protect\right)_{_{\mathit{Purcell's \; \; result}}%}
%\; \approx \; - \; \frac{8}{5} \; \frac{\Omega^3_0 \; \rho \;
%a^2}{ \mu \; Q \; \left[1 \; + \; (c/a)^2 \protect\right]^4} \; 
% (c/a)^4 \; \cos^2 \theta \; \sin \theta
%\label{8.2}
%\ee
One may see that the our calculations reveal that
the inelastic relaxation being much more efficient than predicted in 
Purcell (1979). 
For example, if we define a typical alignment time $\; t_i \;$ as the value of
$\; (d \theta/ dt)^{-1} \;$ at $\; \theta \, = \, \pi/4 \;$ then, for silicate
grains with $\; c/a \; \approx \; 1/2 \;$, ``our'' typical time derived from 
(\ref{8.1}) will be 23 times less than the appropriate estimate\footnote{
Comparing our results with that of Purcell, mind that the value of $\; t_i \;$ 
obtained in (Purcell 1979) was for the degree of suprathermality $\; \beta \, 
= \, 100 \;$.} derived from Purcell's formula. For graphite
grains with $\; c/a \; \approx \; 1/10 \;$, ``our'' typical time will be 
almost 1200 times shorter than the estimate in Purcell (1979). 

III. In reality, the difference between our calculation and the one by Purcell
will be even bigger: due to imperfectness of the grain material, we would 
choose for the $\; Q-$factor lower values than the one suggested by Purcell. 
(See the end of Section VI for discussion.)

\section{Conclusions}

The principal results of this paper are as follows:\\

I. The stresses arising from tumbling of a grain deform it and
 change its moment of inertia, but this change lags behind grain
angular velocity due to non-elastic effects. This process causes
alignment of grain angular momentum with the grain axis of
maximal moment of inertia.

II. Deformations in tumbling grain happen both at the frequency of
precession and at a double frequency. 

III. Inelastic relaxation dominates the Barnett relaxation
for (1) grains with large axis ratio, (2) grains with low $Q$ factor,
i.e. fractal grains produced by coagulation, (3) grains with
$a>2$~$\mu$m, (4) suprathermally rotating grains.

{\bf Acknowledgments}  

The authors are deeply thankful to Eric Heller for his encouragement 
and attention to the work, and to Patrick Thaddeus for moving 
discussions. A.L. would like to acknowledge numerous stimulating 
conversations with Bruce Draine. M.E. wishes to express his thanks to Frans 
Spaepen, Michael Aziz, Martin Bazant and other participants of the Harvard 
Condensed Matter Theory weekly seminar, whose advises on the properties of 
silicates were of an obvious importance for the afore presented study. We also 
want to acknowledge a very useful month-long e-mail conversation on the 
properties of materials, that Michael Ashby kindly and so patiently had with 
us.

\pagebreak

\appendix

\section{Intermediate system of reference\label{A}}

As pointed in Section III, our expression (\ref{28}) for the acceleration of a 
point inside the spinning body considerably differs from an appropriate 
expression presented in the article (Purcell 1979). To be confident in our 
result, we shall now reproduce it by means of a two-step calculation. 

To do so, we shall introduce an ``intermediate'' coordinate system 
($X'$, $Y'$, $Z'$), with basis vectors denoted as $\mathbf{e}$$_{X'}$, 
$\mathbf{e}$$_{Y'}$, and $\mathbf{e}$$_{Z'}$. This system is obtained from 
($X$, $Y$, $Z$) by means of consequent rotations of axis $X$ by the Euler 
angles $\varphi$ and $\theta$. 

In terms of the said angles, 
\begin{equation}
{\Omega}_1 \; \; = \; \; {\dot{\phi}} \; \; \sin \; {\theta} \; \; \sin \; {\psi}
\; \; + \; \; {\dot{\theta}} \; \; \cos \; \psi~~~,
\label{19}
\end{equation}
\begin{equation}
{\Omega}_2 \; \; = \; \; {\dot{\phi}} \; \; \sin \; {\theta} \; \; \cos \; {\psi}
\; \; - \; \; {\dot{\theta}} \; \; \sin \; \psi~~~,
\label{20}
\end{equation}
\begin{equation}
{\Omega}_3 \; \; = \; \; {\dot{\phi}} \; \; \cos \; {\theta}
\; \; + \; \; {\dot{\psi}}~~~,
\label{21}
\end{equation}
and 
\begin{equation}
{{{\Omega}_{\perp}}^{2}} \; \; = \; \; {\Omega}_1^2 \; \; + {\Omega}_2^2 \; \;
=  \; \; {\dot{\phi}}^2 \; {{\sin}^2}{\theta} \; \; + \; \; {\dot{\theta}} ~~~.
\label{22}
\end{equation}
The position, velocity and acceleration of a 
point inside the grain, relative to the ``intermediate'' frame, will be 
denoted by $\mathbf{r'}$, $\mathbf{v'}$ and $\mathbf{a'}$. Evidently,
\begin{equation}
{\bf{r}}'' \; \; = \; \; {\bf{r}}' \; \; = \; \; {\bf{r}} \; \; , \; \; 
\; \; {\bf{e_3}} \; \; = \; \; {\bf{e_{Z'}}} \; \; \; .
\end{equation}
Now, one will be able to calculate the acceleration $\bf{a}$ as 
\begin{equation}
{\bf{a}} \; \; = \; \; {\bf{a'}} \; \; + \; \;
     ({{\bf{e}}_z} \; \ddot{\phi}) \; \times \; {\bf{r'}} \; \; + \; \;
2 \; ({{\bf{e}}_z} \;  \dot{\phi}) \; \times \; {\bf{v'}} \; \; + \; \;
{\dot{\phi}}^2 \; {{\bf{e}}_z} \; \times \;
({{\bf{e}}_z} \; \times \; {\bf{r'}})
\label{29}
\end{equation}
and plug, instead of ${\bf{a'} \;}$ , the expression
\begin{equation}
{\bf{a'}} \; \; = \; \; {\bf{a''}} \; \; + \; \; 
     ({{\bf{e}}_3} \; \ddot{\psi}) \; \times \; {\bf{r''}} \; \; + \; \; 
2 \; ({{\bf{e}}_3} \;  \dot{\psi}) \; \times \; {\bf{v''}} \; \; + \; \;
{\dot{\psi}}^2 \; {{\bf{e}}_3} \; \times \; 
({{\bf{e}}_3} \; \times \; {\bf{r''}})
\label{30} 
\end{equation}
This will once again lead to (\ref{28}).

\section{Diffusion coefficients}

In the absence of gaseous bombardment grain angular momentum stays
constant. The alignment of $\bf J$ in grain axes
is being determined by internal relaxation and that tend to decrease
$\theta$ and thermal fluctuations that randomize $\theta$. To quantify
these processes Fokker-Planck equation may be used.

In the spherical coordinate system $\; J, \theta, \phi \;$, where the polar 
axis is parallel to the principle axis of maximal inertia, the Fokker-Planck 
equation will read: 
\begin{equation}
\frac{\partial f}{\partial t} \; \; = \; \; - \; {\mathbf \nabla} \cdot 
{\mathbf S} \; \; , 
\label{B.1}
\end{equation}
Here $\; f \; = \; f({\bf J}) \;$ is the joint distribution for 
$\; J, \theta \;$ and $\; \phi \;$, while $\; \textbf{S} \;$ is the 
probability current:
\begin{equation}
{\mathbf S} \; \; = \; \; {\mathbf A} \; f \; \; - \; \; \frac{1}{2} 
{\mathbf \nabla}  \cdot (B \; f) \; \; \; ,
\label{B.2}
\end{equation}
where
\begin{equation}
{\mathbf A} \; \; \equiv \; \;  \left\langle
\frac{\Delta {\mathbf J}}{\Delta t} \right \rangle~~~,
\label{b.3}
\end{equation}
is the mean torque, and 
\begin{equation}
B \; \; \equiv \; \; \left\langle
\frac{\Delta {\mathbf J}\Delta {\mathbf J}}{\Delta t}\right \rangle~~~,
\label{b.4}
\end{equation}
where $\langle ..\rangle$ denote ensemble averaging,
is the diffusion tensor.
Generally speaking, $\mathbf A$ and $B$ include the cumulative 
effects of all the processes that change $\; \textbf J \:$ in the body frame.  
Still, in what follows we shall consider the inelastic relaxation as
if it were the sole factor contesting the gas damping. All the other 
mechanisms of orientation will be ignored. 

We have written down the transport equation, and introduced the entities 
$\mathbf A$ and $\mathbf B$, in agreement with (Risken 1984). The difference 
between our definition of $\mathbf A$ and the definition presented in (Landau 
and Lifshitz 1981) stems from the fact that in (Risken 1984) the change of 
momentum of the grain is denoted as $\; {\mathbf{p}} \; + \; {\mathbf{q}} 
\;$, whereas in (Landau and Lifshitz 1981, formula (21.1) and thereabout) it 
is denoted by $\; {\mathbf{p}} \; - \; {\mathbf{q}} \;.$ For this reason 
$\mathbf A$ in formula (21.5) for $\mathbf{S}$, in (Landau and Lifshitz 1981), 
appears with a negative sign.

The inelastic relaxation leaves {\textbf J} unaltered, in the inertial frame. 
In the body frame, which we are using here, the direction of {\textbf J} will 
vary, but its absolute value will be conserved. For this reason the 
components of $\mathbf A$ and $\mathbf B$, having $J$ in their subscripts, 
will vanish. As for $\phi$-dependence, it would be nonexistent should the 
grain be an oblate ellipsoid rather than a prism. Still, in our approximation 
we 
may neglect the $\phi$-dependence: it is, in fact, not that relevant for 
our purposes whether the grain  a circular or a square cross section. 
That is, for the purpose of our estimates (which are anyway not exhaustingly 
exact) we feel free to consider the grain either square, when we need to 
estimate $\; t_{a} \;$ and $\; d{\theta}/dt \;$, or circular, when we need 
to simplify the Fokker-Plank equation. In other words, let us assume that in 
our study of the relaxation over $\theta$ the $\phi$-dependence is to be  
averaged out. Hence all the components of  $\mathbf A$ and $\mathbf B$ with 
$\phi$ in their subscripts will be thrown out either. Now the only
remaining components will be $\; A_{\theta}^{(ir)} \;$ and 
$\; B_{\theta\theta}^{(ir)} \;$ where the extra superscript $\; ^{(ir)} \;$ is 
introduced to emphasize that only the effect of inelastic relaxation (versus 
the gas damping) is taken account of.

The expression for $\; A_{\theta}^{(ir)} \;$ is straightforward from our 
expression for $\; d\theta/dt \;$:
\begin{equation}
A_{\theta}^{(ir)} \; \; =J d\theta/dt 
\label{b.5}
\end{equation}
while $\; B_{\theta\theta}^{(ir)} \;$ is to be found from the principle
of detailed balance: in case the distribution function 
$\; f({\mathbf{J}}) \;$ is
the thermodynamical-equilibrium one ($\; f \; = \; f_{TE} \;$), 
the rate of every microscopic process equals that of its time-reversed
counterpart, and thence the probability current must vanish at
every point in the phase space. In particular,
\begin{equation}
0 \; \; = \; \; S_{\theta} \; \; = \; \; \sin \theta \; \; f_{TE} \; 
A_{\theta}^{(ir)} \; \; - \; \; \frac{1}{2J} \;
\frac{\partial}{\partial\theta} \; \left\{\protect\right. \sin\theta f_{TE} 
B_{\theta \theta}^{(ir)} \left.\protect\right\} 
\label{b.6}
\end{equation}
where the thermodynamical-equilibrium distribution function is a
Boltzmann one\footnote{In the case of a grain being in an
equilibrium, we formally treat it as a small system placed in a bath of 
temperature $T_{\rm grain}.$ This is certainly fair, since the number of  
rotational degrees of freedom of the grain (= 3) is much less than
the amount of its vibrational degrees of freedom (which is of the same order 
as the number of atoms in the grain).}: 
\begin{equation}
f_{TE}({\mathbf{J}}) \; \; = \; \; C \; \exp \left\{\; - \;
E_{rot}(\theta, J)/kT_{\rm grain} \protect\right\}~~~,
\label{b.7}
\end{equation}
C being a normalization constant, $T_{\rm grain}$ being the grain temperature,
and $E_{rot}(\theta, J)$ being the grain rotational energy (expressed by 
equation ($\ref{5.3}$)).
The solution to this first-order differential
equation 
is: 
\begin{equation}
B_{\theta \theta}^{ir} = 
\frac{2J^2\exp(\xi \sin^2\theta)}{t_i \sin\theta} 
\int^{\theta}_{\pi/2} \sin^2y
\left(\tilde{A}\cos^2y + \tilde{B} \sin^2y\right)\exp(-\xi x^2)dx-\frac{J^2
\tilde{A}}{t_i}\exp(\frac{h}{h-1}\xi^2)~~~,
\label{b.8}
\end{equation}
where
\ba
\tilde{A}&=&\frac{2^{3/2} 63 (c/a)^4}{63 (c/a)^4+20}~~~,\\
\tilde{B}&=& \frac{2^{3/2} 20}{63 (c/a)^4+20}
\ea
and
\be
\xi=\frac{(h-1)J^2}{2 I_z k T_{\rm grain}}~~~.
\ee

The solution (\ref{b.8}) is obtained assuming that $B_{\theta \theta}^{ir}
(\pi/2, \xi)=-J^2 \tilde{A}/t_i$, to insure that  $B_{\theta \theta}^{ir}$ is smooth
at $\pi/2$ (compare with Lazarian \& Roberge 1997):
\be
\lim_{\xi\rightarrow \infty}\frac{d^2 B_{\theta \theta}^{ir}(\pi/2, \xi)}
{d\theta^2}<\infty~~~.
\ee 

The coefficients $A_{\theta}$ and $B_{\theta \theta}^{ir}$ can be used to 
describe internal alignment in the presence of gaseous bombardment and
inelastic relaxation as it has been done in Lazarian \& Roberge (1995)
for the case of Barnett relaxation.

\pagebreak

\begin{figure}
\begin{picture}(441,216)
\includegraphics{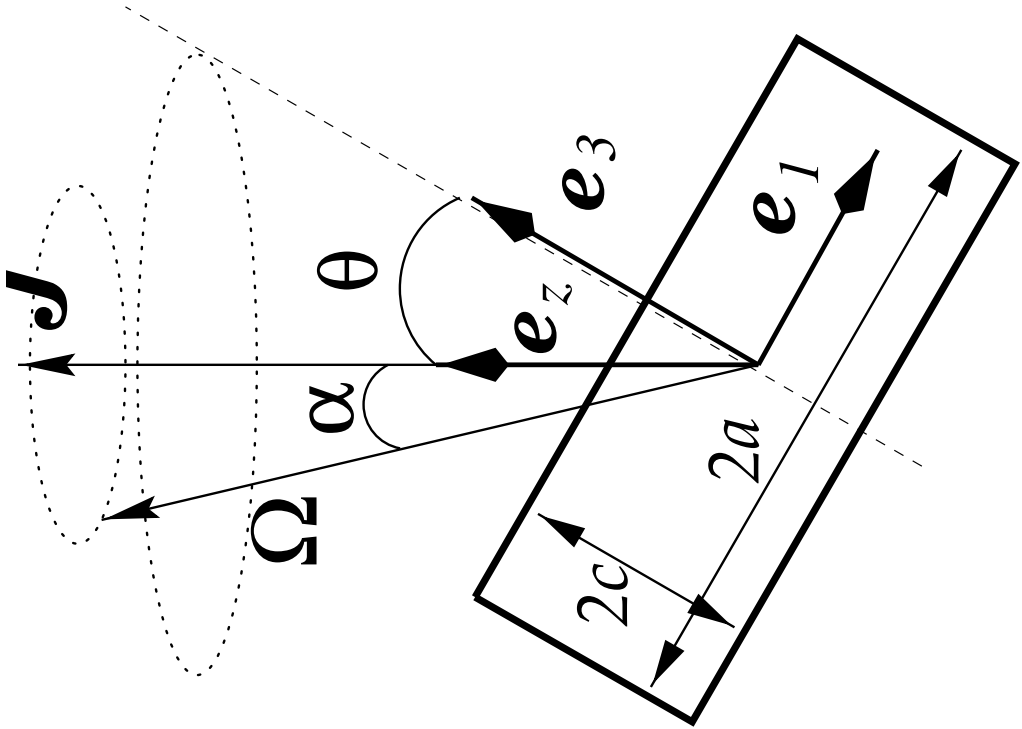}
\end{picture}
\caption[]{The coordinate system $({\bf e}_x, {\bf e}_y, {\bf e}_z)$ is associated with
the inertial frame, so that the (conserved) angular momentum $\bf J$
is aimed along ${\bf e}_z$. The coordinate system  $({\bf e}_1, {\bf e}_2, {\bf e}_3)$ is associated eith the three principal axes of inertia of the body,
so that  ${\bf e}_3$ points along the axis of major inertia.}
\end{figure}

\end{document}